\documentclass[useAMS, referee, usenatbib]{biom}
\def\bSig\mathbf{\Sigma}

\usepackage[T1]{fontenc}        			
\usepackage{cite}  							
\usepackage{amssymb}
\usepackage{amsmath}
\usepackage{bbm}
\usepackage{subfig}
\usepackage{tabularx}
\usepackage{dsfont}
\usepackage{multirow}
\usepackage{lscape}
\usepackage{datetime}
\usepackage{tikz}
\usetikzlibrary{positioning,shapes.geometric}
\newcommand\independent{\protect\mathpalette{\protect\independenT}{\perp}}
\def\independenT#1#2{\mathrel{\rlap{$#1#2$}\mkern2mu{#1#2}}}

\usepackage[utf8x]{inputenc}
\usepackage{longtable}
\usepackage{adjustbox}
\usepackage{xcolor}
\usepackage{bbm}
\usepackage{subfig}
\usepackage{booktabs}

\usepackage[figuresright]{rotating}

\DeclareMathOperator{\E}{\mbox{E}}

\title[Generalizing causal inferences from randomized trials]{Generalizing causal inferences from individuals in randomized trials to all trial-eligible individuals}

\author{Issa J. Dahabreh$^{1,2,3,*}$\email{issa\_dahabreh@brown.edu}, 
Sarah E. Robertson$^{1}$, 
Eric J. Tchetgen Tchetgen$^{4}$, \\
\textbf{Elizabeth A. Stuart$^{\textbf{5}}$, and Miguel A. Hern\'an$^{\textbf{3,6,7}}$ } \\
$^{1}$Center for Evidence Synthesis in Health, Brown University School of Public Health, Providence, RI, U.S.A. \\
$^{2}$Departments of Health Services, Policy \& Practice and Epidemiology, Brown University, Providence, RI, U.S.A. \\
$^{3}$Department of Epidemiology, Harvard-T.H. Chan School of Public Health, Boston, MA, U.S.A. \\
$^{4}$Department of Statistics, Wharton Business School, University of Pennsylvania, PA, U.S.A. \\
$^{5}$Departments of Mental Health, Biostatistics, and Health Policy and Management, \\ Johns Hopkins Bloomberg School of Public Health, Baltimore, MD, U.S.A. \\
$^{6}$Department of Biostatistics, Harvard-T.H. Chan School of Public Health, Boston, MA, U.S.A. \\
$^{7}$Harvard-MIT Division of Health Sciences and Technology, Boston, MA, U.S.A.}

\begin{document}


\date{{\it Received September} 2017. {\it Revised September} 2018.  {\it
Accepted XXX} XXX.}



\pagerange{\pageref{firstpage}--\pageref{lastpage}} 
\volume{XXX}
\pubyear{2018}
\artmonth{XXX}


\doi{XXX}


\label{firstpage}


\begin{abstract}
We consider methods for causal inference in randomized trials nested within cohorts of trial-eligible individuals, including those who are not randomized. We show how baseline covariate data from the entire cohort, and treatment and outcome data only from randomized individuals, can be used to identify potential (counterfactual) outcome means and average treatment effects in the target population of all eligible individuals. We review identifiability conditions, propose estimators, and assess the estimators' finite-sample performance in simulation studies. As an illustration, we apply the estimators in a trial nested within a cohort of trial-eligible individuals to compare coronary artery bypass grafting surgery plus medical therapy versus medical therapy alone for chronic coronary artery disease.
\end{abstract}

%



\maketitle


%

\section{Introduction}
\label{s:intro}

When effect modifiers influence who participates in randomized trials, causal inferences from \emph{randomized individuals} need to be generalized (extended) to \emph{the population of all trial-eligible individuals} \citep{rothwell2005}. The need to extend trial findings arises naturally when trials are nested in cohort studies that collect baseline covariate data from all eligible individuals, including those who are not randomized. In this setting, treatment and follow-up data from non-randomized individuals might be unavailable or unreliable. For example, investigators might wish to use treatment and outcome information only from randomized individuals to avoid confounding of the treatment effect among non-randomized individuals \citep{torgerson1998, silverman1996}. 

In this paper, we build on the emerging literature on ``generalizing'' and ``transporting'' inferences from randomized trials to a target population \citep{cole2010, tipton2012, OMuircheartaigh2014, hartman2013, rudolph2017, Zhang2015} to show how data from randomized trials nested within cohorts of eligible individuals can be used to generalize inferences from randomized individuals to the target population of all trial-eligible individuals. We review identifiability conditions and propose estimators for the potential (counterfactual) outcome means and average treatment effects in the target population. We assess the finite-sample performance of different estimators in simulation studies. Lastly, we illustrate the application of the estimators in the Coronary Artery Surgery Study (CASS), a randomized trial nested within a cohort of trial-eligible individuals to compare coronary artery bypass grafting surgery plus medical therapy versus medical therapy alone for chronic coronary artery disease.

\section{Targets of inference} \label{section_estimands}

Consider a trial nested in a cohort of trial-eligible individuals and let $\mathcal{A}$ be the set of treatments evaluated in the randomized trial. For each treatment $a \in \mathcal{A}$, we use the random variable $Y^a$ to denote the potential (counterfactual) outcome under intervention to receive treatment $a$ \citep{splawaneyman1990, rubin1974, robins2000d}. We only consider a finite number of distinct treatments (e.g., comparisons of treatment vs. control, or comparisons between two or more active treatments).

Baseline covariate information is collected from all cohort members, but treatment and outcome information is only collected (or only deemed reliable) from randomized individuals. We model the data as independent and identically distributed realizations of the random tuple $(X_i, S_i, S_i \times A_i, S_i \times Y_i)$, $i = 1, \ldots, n_{S=1}, n_{S=1}+ 1, \ldots, n_{S=1}+ n_{S=0} = n$, where $n_{S=1}$ is the number of randomized individuals, $n_{S=0}$ is the number of non-randomized individuals, and $n$ is the total number of trial-eligible individuals; $S$ is the indicator for being randomized ($S = 1$ for randomized individuals; $S=0$ for non-randomized trial-eligible individuals); $X$ is a vector of baseline covariates; $A$ is the (randomized) treatment assignment indicator; and $Y$ is the observed outcome. 
An example data structure for binary treatment $A$, along with the potential outcomes, is depicted in Table \ref{table_notation} (throughout, we use uppercase letters to denote random variables and lowercase letters to denote realizations). For simplicity, we assume that full adherence to assigned treatment, absence of measurement error, and no dropout in the trial. Though the methods we propose can be extended to address these issues, we do not pursue these extensions here to simplify exposition and maintain focus on selective trial participation. Extensions of our results will be considered in future work.

\begin{table}[ht]
	\renewcommand{\arraystretch}{1.3}
	\centering
\caption{The left-hand-side of the table depicts the data structure, including baseline covariates $X$, the trial participation indicator ($S = 1$ for the $n_{S=1}$ randomized indviduals and $S=0$ for the $n_{S=0}$ non-randomized individuals), the binary treatment ($A$), and the observed outcome ($Y$). The right hand side of the table depicts the potential outcomes $Y^1$ and $Y^0$ under treatment $a = 1$ and $a = 0$, respectively. The consistency assumption allows us to equate some of these potential outcomes with the observed outcomes, depending on trial participation and treatment assignment. Dashes denote missing values in the observed data.}
	\begin{tabular}{|ccccc||cc|}
		\hline
Individual 		&$X$			&$S$  		&$A$		& $Y$           	    & $Y^1$         		& $Y^0$ \\ \hline\hline
1 				&$x_1$ 			&1  		&1			& $y_1$ 				& $y^1_{1} = y_1$		& $y^0_{1}$  \\ \hline
2 				&$x_2$ 			&1  		&1			& $y_2$     			& $y^1_{2} = y_2$     	& $y^0_{2}$    \\ \hline
$\vdots$ 		&$\vdots$ 		&$\vdots$ 	&$\vdots$	& $\vdots$      		& $\vdots$              & $\vdots$    \\ \hline
$n_1$ 			&$x_{n_1}$		&1			&1 			& $y_{n_1}$     		& $y^1_{n_1} = y_{n_1}$ & $y^0_{n_1}$  \\ \hline\hline 
$n_1 + 1 $ 		&$x_{n_1+1}$  	&1			&0			& $y_{n_1+1}$ 			& $y^1_{n_1+1}$ 		& $y^0_{n_1+1} = y_{n_1+1}$  \\ \hline
$n_1 + 2$ 		&$x_{n_1+2}$ 	&1			&0  		& $y_{n_1 + 2}$ 		& $y^1_{n_1 + 2}$ 		& $y^0_{n_1+2} = y_{n_1 + 2}$ \\ \hline
$\vdots$ 		&$\vdots$    	&$\vdots$   &$\vdots$   		& $\vdots$      		& $\vdots$              & $\vdots$   \\ \hline
$n_1 + n_0 = n_{S=1}$ 	&$x_{n_{\scriptscriptstyle S=1}}$	&1		&0 		& $y_{n_{\scriptscriptstyle S=1}}$ & $y^1_{n_{\scriptscriptstyle S=1}}$ & $y^0_{n_{\scriptscriptstyle S=1}} = y_{n_{\scriptscriptstyle S=1}}$ \\ \hline\hline 
$n_{S=1} + 1$ 		&$x_{n_{\scriptscriptstyle S=1} + 1}$	&0		&$-$ 	& $-$   				& $y^1_{n_{\scriptscriptstyle S=1}+1}$   		& $y^0_{n_{\scriptscriptstyle S=1}+1}$   	 \\ \hline
$n_{S=1} + 2$ 		& $x_{n_{\scriptscriptstyle S=1} + 2}$	&0		&$-$ 	& $-$   				& $y^1_{n_{\scriptscriptstyle S=1} +2}$   		& $y^0_{n_{\scriptscriptstyle S=1}+2}$ \\ \hline
$\vdots$ 		&$\vdots$	&$\vdots$	& $\vdots$  & $\vdots$      		& $\vdots$              & $\vdots$    \\ \hline
$n_{S=1} + n_{S=0} = n$ 			&$x_{n}$		&0			& $-$   	& $-$     & $y^1_{n}$  & $y^0_{n}$           \\ \hline
	\end{tabular}
	\label{table_notation}
\end{table}

We are interested in using the data to draw causal inferences about all trial-eligible individuals. Key targets of inference are the potential outcome means $\E[Y^a]$, for each $ a \in \mathcal A,$ and the average treatment effects $\E [Y^a - Y^{a'}]$, 
for any pair of treatments $ a, a' \in \mathcal{A}.$ In general, $\E [Y^a - Y^{a'}] \neq \E[Y^a - Y^{a'} | S =1];$ in our setup, differences arise when effects are heterogeneous over baseline covariates that are not equidistributed among randomized and non-randomized individuals \citep{dahabreh2016}.

\section{Identifiability conditions for potential outcome means}\label{section_identifiability}

The following conditions are sufficient to identify the potential outcome means $ \E [Y^a ]$ for $a \in \mathcal A $: \emph{(I) Consistency of potential outcomes:} for individuals who receive treatment $A = a$, the observed outcome equals the potential outcome under treatment $a$, that is, if $A_i = a$, then $Y_i = Y^{a}_i,$ for every $a \in \mathcal A.$ \emph{(II) Mean exchangeability in the trial:} $ \E [Y^a | X, S=1, A=a]= \E [Y^a | X, S=1]$ for every $a \in \mathcal A.$ \emph{(III) Positivity of treatment assignment probability in the trial:} $\Pr[A=a | X = x, S=1] > 0$ for every $a \in \mathcal A$ and every $x$ such that $f_{X|S}(x | S = 1) > 0.$ \emph{(IV) Mean generalizability (i.e., mean exchangeability over $S$):} $ \E[Y^a| X, S=1] = \E[Y^a |X]$ for every $a \in \mathcal A.$ The mean generalizability condition is implied by, but does not imply, the often invoked (and stronger) condition of generalizability in distribution, $Y^a \independent  S | X.$ \emph{(V) Positivity of trial participation:} $ \Pr[S=1 | X = x] >0$ for every  $x$ such that $f_X(x) > 0$.

Conditions $I$ through $III$ are expected to hold for well-defined interventions compared in randomized trials \citep{Hernan2015}. 
Of note, implicit in our notation is an assumption that the invitation to participate in the trial and trial participation itself do not affect the outcome except through treatment assignment -- this assumption is often plausible in pragmatic randomized trials \citep{Ford2016}. Conditions $IV$ and $V$ are needed to extend inferences about the potential outcome means (or the average treatment effect) from randomized individuals to the target population of all trial-eligible individuals.


As is common in the causal inference literature, we use the term ``exchangeability'' to mean that two or more groups are expected to have the same outcome functionals had they received the same treatment \citep{greenland1986}. This notion differs from other uses of the term in statistics. Furthermore, in this section, we used the term ``consistency'' for the condition linking potential (counterfactual) and observed outcomes; later, we will use the same term to refer to the property of an estimator that converges to its estimand; the intended meaning should be clear from context.

\section{Identification}\label{section_identification}

Under the identifiability conditions listed in the previous section, the conditional potential outcome mean in the target population of all trial eligible individuals for each treatment $a \in \mathcal A$ can be identified,
\begin{equation*}
		\E [Y^a | X ] = \E [Y^a | X, S = 1] =  \E [Y^a | X, S = 1, A = a] =  \E [Y | X, S = 1, A = a].
\end{equation*}
It follows that we can identify the potential outcome means,
\begin{equation*}
		\E [Y^a] = \E_{X}\! \big[ \E [Y | X, S = 1, A = a]\big], \mbox{ for each } a \in \mathcal A,
\end{equation*}
where the subscript $X$ denotes expectation with respect to the target population distribution of $X$. Using this result, we can identify average treatment effects by differencing, $$ \E [Y^a - Y^{a^\prime}] = \E_{X}\! \big[ \E [Y | X, S = 1, A = a]\big] -\E_{X}\! \big[ \E [Y | X, S = 1, A = a^{\prime}]\big].$$ Furthermore, we can identify any other measure of effect defined in terms of the potential outcome means \citep{Hernan2004}; for example, for binary $Y$, we can identify the causal risk ratio comparing treatments $a$ and $a^\prime$ in the target population, $$ \dfrac{ \Pr [Y^a = 1]}{ \Pr [Y^{a^\prime} = 1]} = \dfrac{ \E [Y^a]}{ \E [Y^{a^\prime} ]} = 
\dfrac{ \E_{X}\! \big[ \E [Y | X, S = 1, A = a]\big] }{ \E_{X}\! \big[ \E [Y | X, S = 1, A = a^\prime]\big]}.$$

In the Appendix, we show that identification of average treatment effects is possible under weaker identifiability conditions, which do not, however, suffice to identify the potential outcome means. Because these potential outcome means are of inherent scientific and policy interest in most applications, in the remainder of the paper we focus on estimating them.

\section{Estimation} \label{section_estimators}

We now discuss estimators of the functional $\mu(a) \equiv \E_{X}\! \big[ \E [Y | X, S = 1, A = a]\big].$ Specifically, we consider (1) \emph{outcome model-based} estimators that rely on modeling the expectation of the outcome; (2) \emph{inverse probability (IP) weighting} estimators that rely on modeling the probability of participation in the trial; and (3) \emph{augmented IP weighting} estimators that rely on modeling both the expectation of the outcome and the probability of participation in the trial. Hereafter, ``convergence'' and the symbol ``$\rightarrow$'' denote convergence in probability; estimators that converge to their corresponding estimands are termed ``consistent'' (see the Web Appendix for additional information). 

\subsection{Outcome model-based estimator}\label{outcome_model}


The outcome model-based estimator is obtained by fitting a conditional outcome mean model among trial participants and marginilizing over the empirical covariate distribution of all trial-eligible individuals \citep{robins1986},
\begin{equation}\label{estimator_OR}
\widehat\mu_{\text{\tiny OM}} (a) = \dfrac{1}{n} \sum \limits_{i = 1}^{n}  \widehat g_{a}(X_i) ,
\end{equation}
where $\widehat g_{a}(X)$ is an estimator of $ \E [Y | X, S = 1, A = a ]$ for $a \in \mathcal A$. Typically, we posit a parametric outcome model $g_{a}(X; \theta)$ with finite-dimensional parameter $\theta$ for each treatment $a$, and estimate the model parameters by standard methods. In applications, we recommend fitting separate outcome models among the treated and untreated randomized individuals to better capture effect modification over baseline covariates. When the outcome models are correctly specified, $ g_{a}(X; \widehat{\theta}) \rightarrow \E [Y | X, S = 1, A = a ]$ and $\widehat\mu_{\text{\tiny OM}} (a) \rightarrow \mu(a)$ for each $ a \in \mathcal A$.

\subsection{IP weighting estimators}


An alternative approach for generalizing inferences from randomized individuals to the target population relies on estimating the probability of trial participation followed by IP weighting, an approach related to the analysis of sampling surveys \citep{horvitz1952} and to IP weighting methods for confounding control in observational studies \citep{robins1999association}. Specifically, we can use the IP weighting estimator 
\begin{equation} \label{estimator_IPW1}
\widehat\mu_{\text{\tiny IPW1}}(a)  = \frac{1}{n} \sum_{i = 1}^{n} \dfrac{S_i I (A_i = a)  Y_i}{\widehat p(X_i) \widehat e_a(X_i),},
\end{equation}
where $\widehat p(X)$ is an estimator for $\Pr[S = 1 | X] $ and $\widehat e_a(X)$ is an estimator for $\Pr[A = a | X, S = 1]$ for $a \in \mathcal A.$ When trials are nested within cohorts of eligible individuals, we do not know the probability of trial participation, but we can estimate it, typically using a parametric model $p(X; \beta)$. The probability of treatment in the trial $\text{Pr}[A=a | X, S = 1 ]$ is under the control of the investigators and does not need to be estimated. Nevertheless, estimating the probability of treatment among randomized individuals, say using a parametric model $e_a(X; \gamma)$, can improve efficiency in finite samples. Heuristically, modeling the probability of treatment is beneficial because it addresses random imbalances in baseline covariates among randomized individuals, provided that we properly account for the estimation of $\gamma$ \citep{hahn1998role, Lunceford2004, williamson2014variance}. When the model for the probability of participation is correctly specified, and given that the model for the probability of treatment among randomized individuals is always correctly specified, $p(X; \widehat \beta) e_a(X; \widehat \gamma) \rightarrow \text{Pr}[S = 1 | X ] \text{Pr}[A=a | X, S = 1 ]$ and $\widehat\mu_{\text{\tiny IPW1}} (a) \rightarrow  \mu(a)$ for every $a \in \mathcal A$. 

The estimator in (\ref{estimator_IPW1}) is unbounded, in the sense that it can produce point estimates that fall outside the support of $Y$, particularly in the presence of extreme weights \citep{robins2007}. Using the identity 
\begin{equation}\label{eq:weight_identity}
 \E \left[ \dfrac{S I(A = a) }{\Pr[S = 1 | X] \Pr[A = a | X, S = 1]} \right] = 1,
\end{equation} 
which holds when positivity conditions $III$ and $V$ hold, we can construct a bounded IP weighting estimator by normalizing the weights to sum to 1 \citep{hajek1971comment},
\begin{equation} \label{estimator_IPW2}
\widehat\mu_{\text{\tiny IPW2}} (a) = \left\{ \sum_{i = 1}^{n} \dfrac{S_i I (A_i = a) }{\widehat p(X_i) \widehat e_a(X_i)} \right\}^{-1} \sum_{i = 1}^{n} \dfrac{S_i I (A_i = a)  Y_i}{\widehat p(X_i) \widehat e_a(X_i)}.
\end{equation}

Even when not intending to use IP weighting estimators, analysts should inspect the distribution of the estimated probabilities of trial participation to examine overlap between randomized and non-randomized individuals. Furthermore, analysts should examine whether the sample analog of (\ref{eq:weight_identity}) is approximately satisfied, that is, $$ \dfrac{1}{n} \sum\limits_{i=1}^{n} \dfrac{S_i I(A_i = a) }{\widehat p(X_i) \widehat e_a(X_i)} \approx 1.$$

\subsection{Augmented IP weighting estimators}

We can combine modeling the probability of trial participation with modeling the expectation of the outcome among randomized individuals to obtain more efficient (augmented) IP weighting estimators \citep{Robins1994, Bang2005}. Augmented IP weighting estimators are also \emph{doubly robust} in the sense that they are consistent and asymptotically normal when either the model for the probability of participation or the model for the expectation of the outcome is correctly specified. Because background knowledge is often inadequate to correctly specify the outcome model, the improved efficiency of augmented IP weighting estimators is often the primary motivation for using them \citep{Tan2007}, provided the model for the probability of trial participation can be (approximately) correctly specified.

The theory of augmented IP weighting estimation is extensive and multiple estimators, with different behavior in finite samples, are doubly robust \citep{Robins1994, robins2007}. Here, we examine three estimators that are easy to implement in standard statistical packages.

We begin by considering the augmented IP weighting estimator
\begin{align}\label{estimator_AIPW1}
\begin{split}
\widehat{\mu}_{\text{\tiny AIPW1}} (a)	 =  \frac{1}{n} \sum_{i=1}^{n} \left\{   \frac{S_{i}  I (A_{i} = a) }{\widehat p(X_i) \widehat e_a(X_i)} \big\{ Y_{i} - \widehat g_a (X_i) \big\} + \widehat g_a (X_i)   \right\},
\end{split}
\end{align}
with $\widehat p(X)$, $\widehat e_a(X)$, and $\widehat g_a (X)$ as defined above.

We can normalize the weights, as we did for the IP weighting estimators, to obtain 
\begin{align}\label{estimator_AIPW2}
\begin{split}
\widehat{\mu}_{\text{\tiny AIPW2}} (a) =  \left\{ \sum_{i = 1}^{n} \dfrac{S_i I (A_i = a) }{\widehat p(X_i) \widehat e_a(X_i)} \right\}^{-1} \sum_{i=1}^{n}    \frac{S_{i}  I (A_{i} = a) }{\widehat p(X_i) \widehat e_a(X_i)} \big\{ Y_{i} - \widehat g_a (X_i) \big\}   +  \dfrac{1}{n}   \sum_{i = 1}^{n} \widehat g_a (X_i).
\end{split}
\end{align}

Alternatively, we can obtain a bounded regression-based augmented IP weighting estimator by fitting an IP weighted parametric multi-variable regression model for the outcome among randomized individuals and then standardizing the predictions over the empirical covariate distribution of all trial-eligible individuals,
\begin{equation}\label{estimator_AIPW3}
\widehat\mu_{\text{\tiny AIPW3}} (a) = \frac{1}{n} \sum_{i = 1}^{n} g_a (X_i; \widetilde\theta),
\end{equation}
where  $\widetilde\theta$ is the vector of estimated parameters from the IP weighted outcome regression. This estimator is doubly robust when the outcome is modeled with a linear exponential family quasi-likelihood \citep{gourieroux1984} and the canonical link function \citep{robins2007, wooldridge2007}.

In the Web Appendix we show that the estimator in (\ref{estimator_AIPW1}) is the one-step in-sample estimator suggested by the influence function for $\mu(a)$ and argue that it is locally efficient \citep{Robins1994, Robins1995}. The estimators in (\ref{estimator_AIPW2}) and (\ref{estimator_AIPW3}) are asymptotically equivalent to the estimator in (\ref{estimator_AIPW1}). In finite samples, augmented IP weighting estimators will tend to produce more precise results than non-augmented IP weighting estimators, sometimes strikingly so. When using any augmented IP weighting estimator, as for the outcome model-based estimator in (\ref{estimator_OR}), we recommend fitting separate outcome models in each treatment group in the randomized trial.

\section{Simulation studies} \label{section_simulations}

We conducted simulation studies to compare the finite-sample performance of different estimators, for binary and continuous outcomes. Details about the simulation study methods and code to replicate the analyses are provided in the Web Appendix; the simulation results are summarized in Appendix Tables A.2 through A.19. In brief, our numerical studies confirmed that, when all models were correctly specified, all estimators were approximately unbiased even with small numbers of randomized individuals and small total cohort sample sizes (when one model was incorrectly specified, the augmented IP weighting estimators were also approximately unbiased; results not shown). The outcome-model based estimator had the lowest variance, followed closely by the two doubly robust estimators; IP weighting estimators had substantially larger variance than all other estimators, especially when using non-normalized weights.

\section{The Coronary Artery Surgery Study (CASS)} \label{section_example}

\subsection{CASS design and data}

CASS was a comprehensive cohort study that compared coronary artery bypass grafting surgery plus medical therapy (henceforth, ``surgery'') versus medical therapy alone for individuals with chronic coronary artery disease; details about the design of CASS are available elsewhere \citep{william1983, investigators1984}. In brief, individuals undergoing angiography in 11 participating institutions were screened for eligibility and the 2099 trial-eligible individuals who met the study criteria were either randomized to surgery or medical therapy (780 individuals), or included in an observational study (1319 individuals). We excluded 6 individuals for consistency with prior CASS analyses and in accordance with CASS data release notes; in total we used data from 2093 individuals (778 randomized; 1315 non-randomized). Baseline covariates were collected from randomized and non-randomized individuals in an identical manner. No randomized individuals were lost to follow-up in the first 10 years of the study; we did not use information on adherence among randomized individuals, in effect assuming that the non-adherence would be similar among all eligible individuals, so that intention-to-treat effects are transportable.

\subsection{Statistical analysis}

\paragraph{Estimands} We estimated the 10-year mortality risk under surgery and medical therapy, and the risk difference and risk ratio comparing the treatments for the target population of all trial-eligible individuals.

\paragraph{Model specification} We fit logistic regression models for the probability of participation in the trial, the probability of treatment among randomized individuals, and the probability of the outcome (in each treatment arm) with the following covariates: age, severity of angina, history of previous myocardial infarction, percent obstruction of the proximal left anterior descending artery, left ventricular wall motion score, number of diseased vessels, and ejection fraction. We chose these variables based on a previous analysis of the same data \citep{olschewski1992} and did not perform any model selection.

\paragraph{Missing baseline covariate data} Of the 2093 trial-eligible individuals, 1686 had complete data on all baseline covariates (731 randomized, 368 in the surgery group and 363 in the medical therapy group; 955 non-randomized). For simplicity, in the main text we only report analyses restricted to individuals with complete data. To examine whether missing data influenced our results, we repeated our analyses using (1) multiple imputation with different models for the missing data conditional on the observed data (multivariate normal multiple imputation and multiple imputation with chained equations) \citep{little_rubin_2014statistical} and (2) IP of missingness weighting for non-monotone missing data \citep{sun2018inverse}. 

\paragraph{Inference} For all analyses, we used bootstrap resampling (with 10,000 samples) to obtain normal distribution-based 95\% confidence intervals (results using percentile intervals were very similar and are not shown).

\subsection{Results}

\begin{figure}[ht!]
	\centering
	\caption{Kernel densities of the estimated probabilities of trial participation (left panel) for randomized (solid line) and non-randomized (dashed line) individuals and estimated weights (right panel) for randomized individuals.}
	\includegraphics[scale=0.9]{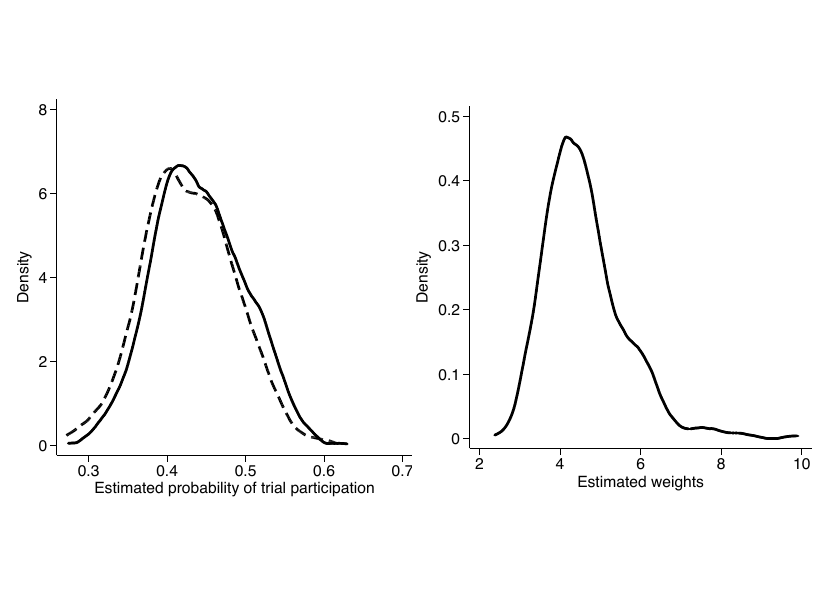}
	\label{fig:probs_and_weights}
\end{figure}

\paragraph{Baseline characteristics, overlap, and balance} Web Appendix Table A.20 summarizes baseline covariate information for randomized and non-randomized individuals. The left panel of Figure \ref{fig:probs_and_weights} presents kernel densities of the estimated probabilities of trial participation for randomized and non-randomized individuals; there was good overlap and the smallest estimated probability for trial participation for randomized individuals was approximately 0.273. The right panel of Figure \ref{fig:probs_and_weights} presents the kernel density of the estimated weights among randomized individuals, obtained as the inverse of the estimated probability of trial participation times the inverse of the estimated probability of receiving the treatment actually received among randomized individuals. The sample average of the estimated weights was approximately 1.001, both in the surgery and medical therapy groups; the largest estimated weight was less than 10. Taken together, these results suggest that the observed covariate distributions of randomized and non-randomized individuals had sufficient overlap for attempting to extend inferences from the trial to the target population of all eligible individuals. Baseline covariates in randomized and non-randomized individuals were balanced after IP weighting (Web Appendix Table A.21).

\paragraph{Treatment-specific risks and treatment effects} Estimates of the 10-year mortality risk and treatment effects at 10 years are shown in Table \ref{table_example}. All generalizability methods produced similar results: the mortality risk in the target population was about 18\% for surgery and 20\% for medical therapy, corresponding to a risk difference of about 2\% and a risk ratio of about 0.9, in favor of surgery. Because different methods rely on different parametric models, agreement across their point estimates suggests that inference is not driven by model specification. The similar confidence interval widths produced by different generalizability estimators reflect the binary nature of the outcome as well as the absence of strong selection on measured baseline covariates in this particular application. The mortality risk among randomized individuals was about 17.3\% for surgery and 20.9\% for medical therapy, corresponding to a risk difference of about 3.5\% in favor of surgery. The closeness of estimates from the randomized trial and our generalizability analyses probably reflects the absence of strong selection on measured baseline covariates in this application.

\begin{sidewaystable}[ht]
\centering
\caption{Estimated 10-year mortality risk (\%) for surgery and medical therapy, and risk difference and risk ratio comparing the treatments among randomized individuals and all trial-eligible individuals in the Coronary Artery Surgery Study. For each point estimate we provide bootstrap confidence intervals from 10,000 re-samplings. Surgery = coronary artery bypass grafting surgery plus medical therapy; CI = confidence interval; Medical = medical therapy; RD = risk difference; RR = risk ratio; Trial-only = unadjusted analysis using only observations from randomized individuals; OM = outcome model-based standardization; IPW1 = inverse probability weighting; IPW2 = inverse probability weighting with normalized weights; AIPW1 = augmented inverse probability weighting; AIPW2 = augmented inverse probability weighting with normalized weights; AIPW3 = inverse probability weighted regression; (eq.) gives the equation number for each estimator.}
\label{table_example}
{\large
\begin{tabular}{llcccccccc}
\hline
\textbf{Target population}                                                             & \textbf{Estimator (eq.)} & \textbf{Surgery (\%)} & \textbf{95\%  CI} & \textbf{Medical (\%)} & \textbf{95\%  CI} & \textbf{RD (\%)} & \textbf{95\%  CI} & \textbf{RR} & \textbf{95\%  CI} \\ \hline
Randomized individuals                                                                     & Trial-only         & 17.4            & (13.5 , 21.3 ) & 20.1            & (15.9 , 24.3 ) & -2.7       & (-8.5, 3.0)  & 0.86        & (0.59 , 1.14)     \\ \hline
\multirow{6}{*}{\begin{tabular}[c]{@{}l@{}}All trial-eligible\\ individuals\end{tabular}} & OM (\ref{estimator_OR})    & 18.5            & (14.3 , 22.6) & 20.0            & (15.9 , 24.0) & -1.5       & (-7.2 , 4.2)  & 0.93        & (0.64, 1.21)     \\
& IPW1 (\ref{estimator_IPW1})        & 18.0            & (13.9 , 22.1) & 19.9            & (15.9 , 24.0) & -2.0       & (-7.7 , 3.8)  & 0.90        & (0.62, 1.18)     \\
& IPW2 (\ref{estimator_IPW2}) & 18.0            & (13.9 , 22.0) & 19.9            & (15.9 , 24.0) & -2.0       & (-7.7 , 3.8)  & 0.90        & (0.62, 1.18)     \\
& AIPW1 (\ref{estimator_AIPW1}) & 18.3            & (14.2 , 22.4) & 20.0            & (15.9 , 24.0) & -1.6       & (-7.3 , 4.1)  & 0.92        & (0.64, 1.20)     \\
& AIPW2 (\ref{estimator_AIPW2}) & 18.3            & (14.2 , 22.4) & 20.0            & (15.9 , 24.0) & -1.6       & (-7.3 , 4.1)  & 0.92        & (0.64, 1.20)     \\
& AIPW3 (\ref{estimator_AIPW3})   & 18.3            & (14.2 , 22.4) & 19.9            & (15.9 , 23.9) & -1.6       & (-7.2 , 4.1)  & 0.92        & (0.64, 1.20)     \\ \hline
\end{tabular}
}
\end{sidewaystable}

\paragraph{Using treatment and outcome data on non-randomized individuals to evaluate the generalizability analyses} In CASS, data on treatment and outcome were collected among non-randomized individuals, even though such data are not necessary for identification under the conditions in Section \ref{section_identifiability}. To evaluate whether outcome models built among randomized individuals were reasonable, we compared the empirical mortality risk among non-randomized individuals who received treatment $a$ against the average outcome predictions for the same group of individuals, using models estimated among randomized individuals. The similarity of the empirical risk in non-randomized individuals to the average of the predictions provides some reassurance that the models were not grossly inappropriate for generalizing trial results: for surgery, the empirical risk was $18.6\%$ and the average of the predictions was $19.8\%$; for medical therapy, the empirical risk was $19.1\%$ and the average of the predictions was $18.9\%$. 

The comparison in the preceding paragraph is only indirectly relevant to the goal of generalizing inferences to the target population of all trial-eligible individuals; after all, the comparison is conditional on treatment received among non-randomized individuals. Another way to evaluate the generalizability analyses, is to compare them against an observational analysis that uses all the data (regardless of $S$). The validity of such observational analysis requires exchangeability of treatment groups, $\E[Y^a | X, S, A = a] = \E[Y^a | X, S]$, and positivity, $\Pr[A = a | X, S] > 0$, for $a \in \{0,1\}$, but does not require identifiability conditions $IV$ or $V$ (which are necessary to endow the generalizability estimators with a causal interpretation). Using an IP weighting regression estimator for confounding control in the entire CASS dataset \citep{robins2007}, the potential outcome mean estimates were $17.9\%$ and $19.6\%$ for surgery and medical therapy, respectively; these estimates are similar to those from the generalizability analyses. Because the observational and generalizability analyses rely on different identifiability conditions, agreement between them provides some mutual support (but does not establish the validity of either).

\paragraph{Comparison of methods for handling missing data} Web Appendix A.22 presents descriptive statistics for missing covariate data; the missing data pattern for baseline covariates was non-monotone. The point estimates from analyses using multiple imputation or IP weighting for missing baseline covariates were nearly identical to those of the complete case analyses and only slightly more precise (Web Appendix Tables A.23 through A.25).

\section{Discussion}

We examined methods for extending inferences from randomized individuals to the target population of trial-eligible individuals, using trials nested within cohorts for outcomes observed at a single time-point post-randomization. Our work adds to the literature on ``generalizability'' and ``transportability'' \citep{cole2010,OMuircheartaigh2014, tipton2012, hartman2013, Zhang2015,rudolph2017}, and connects with the literature on selection bias \citep{keiding2016perils, infante2018reflection}. 

The methods we propose rest on two identifiability conditions beyond those supported by randomization: \emph{mean generalizability} from randomized to non-randomized individuals and \emph{positivity of trial participation}. Directed acyclic graphs can facilitate reasoning about the mean generalizability condition \citep{pearl2015, pearl2014}. Arguably, the identifiability conditions are most plausible in studies explicitly designed to collect information on all trial-eligible individuals, including those who are not randomized. Our methods are therefore appropriate for ``comprehensive cohort studies'' \citep{olschewski1985, olschewski1992, schmoor1996}, randomized preference designs \citep{torgerson1998, lambert2000}, and pragmatic trials embedded within healthcare systems \citep{fiore2016, choudhry2017}. Though our approach is useful for extending inferences from randomized individuals to trial-eligible individuals in these designs, it does not address individuals who are candidates for treatment but do not meet the trial eligibility criteria.

Our approach uses treatment and outcome data only from randomized individuals, to avoid confounding of the treatment effect among non-randomized individuals and eliminate the need to follow them up. Nevertheless, if treatment and outcome data from non-randomized individuals are available, analysts can use them to assess model specification, as we did in the CASS reanalysis. When such data are available and substantive knowledge suggests that the observed covariates suffice to adjust for confounding among non-randomized individuals, it may be useful to compare the results of generalizability methods against the results of observational analyses of the entire cohort. These analyses target the same causal quantities but rest on different identifiability conditions: generalizability analyses require mean generalizability from randomized to non-randomized individuals whereas observational analyses require mean exchangeability of treated and untreated individuals. As in our CASS reanalysis, similarity of results from analyses that rest on different identifiability conditions provides mutual support for their validity (but does not definitively establish it).

Methods for generalizing inferences from randomized individuals to all trial-eligible individuals exploit models for the expectation of the outcome in the trial or the probability of trial participation. Investigators typically rely on parametric working models and model misspecification can lead to inconsistency, even if mean generalizability holds. Approximately correct specification of the model for the probability of participation may be more feasible, because inverstigators can use surveys or qualitative studies to investigate what drives eligible individuals to participate in a randomized trial \citep{ross1999}. The findings from these investigations can be used to specify models for trial participation and obtain consistent IP weighting estimators. 

Augmented IP weighting estimators are consistent when either working model is correctly specified, providing two opportunities for valid inference. Misspecification of just one of the models can adversely affect how these estimators perform. And misspecification of both models, in some cases \citep{waernbaum2017model}, can make augmented IP weighting estimators perform worse than the (misspecified) outcome model-based estimator \citep{kang2007}. Serious bias can also occur when IP weights are highly variable, though this problem is to some extent mitigated with bounded estimators \citep{robins2007}. Even when the outcome model is misspecified, augmented IP weighting estimators may still be preferred for their typically increased efficiency compared to non-augmented IP weighting estimators; their performance may be further improved with approaches that explicitly attempt to minimize variance \citep{cao2009improving, rotnitzky2012improved} or reduce bias \citep{vermeulen2015bias} under model misspecification.

\section{Supplementary Materials}

Web Appendices, Tables, and Figures referenced in Sections \ref{section_estimators} through \ref{section_example}, including example \texttt{R} code implementing the methods, are available with this paper at the Biometrics website on Wiley Online Library.


\backmatter


\section*{Acknowledgments}

This work was supported by Patient-Centered Outcomes Research Institute (PCORI) awards ME-1306­-03758 and ME­-1502­-27794 and National Institutes of Health (NIH) grant R37 AI102634. Statements in this paper do not necessarily represent the views of the PCORI, its Board of Governors, the Methodology Committee, or the NIH. The data analyses in our paper used CASS research materials obtained from the NHLBI Biologic Specimen and Data Repository Information Coordinating Center. This paper does not necessarily reflect the opinions or views of the CASS or the NHLBI. The authors thank Christopher H. Schmid and Bora Youn (both at Brown University) for helpful discussions.



%

\bibliographystyle{biom} 
\bibliography{biometrics_format}

\label{lastpage}

\end{document}